# Phase composition and transformations in magnetron-sputtered (Al,V)$_2$O$_3$ coatings

L. Landälv[*,a,b], C-F. Carlström[b], J. Lu[a], D. Primetzhofer[c], M. J. Jõesaar[d], M. Ahlgren[b], E. Göthelid[b], B. Alling[e], L. Hultman[a], P. Eklund[*,a]

[a]Thin Film Physics Division, Department of Physics, Chemistry, and Biology (IFM), Linköping University, Linköping SE-581 83, Sweden
[b]Sandvik Coromant AB, Stockholm SE-12680, Sweden
[c]Applied Nuclear Physics, Department of Physics and Astronomy, Uppsala University, Uppsala SE-75120, Sweden
[d] SECO Tools AB, SE-737 82 Fagersta, Sweden
[e]Theoretical physics, Department of Physics, Chemistry, and Biology (IFM), Linköping University, Linköping SE-581 83, Sweden

**Abstract**

Coatings of (Al$_{1-x}$V$_x$)$_2$O$_3$, with x ranging from 0 to 1, were deposited by pulsed DC reactive sputter deposition on Si(100) at a temperature of 550 °C. XRD showed three different crystal structures depending on V-metal fraction in the coating: α-V$_2$O$_3$ rhombohedral structure for 100 at.% V, a defect spinel structure for the intermediate region, 63 - 42 at.% V. At lower V-content, 18 and 7 at.%, a gamma-alumina-like solid solution was observed, shifted to larger d-spacing compared to pure γ-Al$_2$O$_3$. The microstructure changes from large columnar faceted grains for α-V$_2$O$_3$ to smaller equiaxed grains when lowering the vanadium content toward pure γ-Al$_2$O$_3$. Annealing in air resulted in formation of V$_2$O$_5$ crystals on the surface of the coating after annealing to 500 °C for 42 at.% V and 700 °C for 18 at.% V metal fraction respectively. The highest thermal stability was shown for pure γ-Al$_2$O$_3$-coating, which transformed to α-Al$_2$O$_3$ after annealing to 1100° C. Highest hardness was observed for the Al-rich oxides, ~24 GPa. The latter decreased with increasing V-content, larger than 7 at.% V metal fraction. The measured hardness after annealing in air decreased in conjunction with the onset of further oxidation of the coatings.

∗Corresponding authors
Email address: ludvig.landalv@liu.se (L. Landälv)
Per.eklund@liu.se (P. Eklund)

# 1. Introduction

Aluminum oxide is an important material because of its chemical inertness, high melting point, high hardness and high electrical resistance [1, 2]. It is therefore used in numerous applications such as diffusion barriers [3], dielectric material [4], refractory material and abrasive or cutting tool material [1, 5].

In metal cutting applications, chemical-vapor-deposited (CVD) α-$Al_2O_3$ has been the dominating tool coating for several decades, mainly due to its excellent wear properties compared to other coating materials [6, 7]. CVD oxides are typically used as wear protection coatings in rough metal machining, e.g., turning or milling applications [8]. Over the last ~15 years, physical vapor deposition (PVD) of $Al_2O_3$ based oxide coatings have been introduced with the aim to broaden the use of oxide coatings in metal cutting [9, 10]. PVD oxides offer possibilities for use in finishing and solid round tool applications [11, 12] as well as potentially enable the use of wear-resistant oxides on temperature-sensitive substrates [13]. However, due to the relatively low growth temperatures used in PVD, other metastable $Al_2O_3$ phases such as γ, θ, and κ are often obtained [10, 13, 14]. Various PVD synthesis routes have been evaluated for promoting α-phase formation at low deposition temperatures such as ion bombardment [15], the use of crystallographic templates [16], and alloying of, e.g., $Cr_2O_3$ in $(Al,Cr)_2O_3$ [9, 17]. The latter enable solid-solution stabilization, since $Cr_2O_3$ crystallizes exclusively in the corundum structure and has similar lattice constants to α-$Al_2O_3$.

In analogy to phase stabilization of $(Al,Cr)_2O_3$, vanadium oxide $V_2O_3$ also crystallizes in the corundum structure, which may suggest that V alloying promotes stabilization of corundum structure $(Al,V)_2O_3$. Vanadium exist in several oxidation states, which results in several possible oxides with varying properties. From low to high oxidation state, the most common phases are:

$V_2O_3$ which is interesting for its magnetic properties (presumably Mott transition) [18, 19], $V_3O_5$ Magnéli series oxide, [20] showing interesting magnetic properties with the highest metal to insulator switching temperature of the series [21, 22], $VO_2$, as primarily investigated for its thermochromic property, i.e., for architectural glass, [23-25] and $V_2O_5$, which shows catalytic properties and good ionic conduction [26] as well as low friction due to low melting temperature ~700 °C [27]. The O-V phase diagram summarizes these phases [28]. Calculated mechanical properties are surveyed by Reeswinkel et al. [29]. Many of the reported vanadia and aluminum vanadium oxide phases have been produced through different melting processes, precipitation or reduction of powders, i.e., rhombohedral $V_2O_3$ with different amount of Mo [30], triclinic $AlVO_4$ [31], cubic spinel $AlVO_3$ [32], $AlVO_4$, and orthorhombic $V_2O_5$ [33]. In addition to these studies, liquid phase ($V_2O_5$) sintering of $Al_2O_3$ decreased the α-phase formation temperature in conjunction with enhanced grain growth [34, 35]. The process routes used to form theses pseudobinary $Me_xV_yO_z$ compounds are controlled by thermodynamic equilibrium processes and the use of intrinsically non-equilibrium process techniques, such as PVD, may promote other phases useful for the cutting tool industry.

So far V-based oxides, in the context of cutting tools, earned interest based on its possible low friction such as for the layered Magnéli phases [36] and low-temperature melting (~680°C) of $V_2O_5$ [37]. Much work has its starting point in PVD-deposited TiAlN, adding V into the coatings, in order to explore the effect of formation of the different vanadium oxides upon oxidation studies or tribo-testing at elevated temperatures. The V-containing low-friction nitride approach is summarized in a review by Franz et al. [38] and shows promising low-friction effects up to 600-700 °C. The diffusion behavior in a nitride is, however, different from an oxide, and the direct synthesis of the AlVO oxide with PVD, for a vide set of Al-V composition remains to be investigated, for the aim to determine phase formation.

In the present work, we therefore investigate PVD Al-V-O coatings along the $Al_2O_3$-$V_2O_3$ compositional line, probing for possible α-structured oxide solid solutions, and their temperature stability. $(Al_{1-x}V_x)_2O_3$, $0 \leq x \leq 1$ coatings have been synthesized by reactive magnetron sputtering followed by a high temperature annealing in air. Corresponding phase and crystal structure of the as-formed coatings with respect to their mechanical properties are also analyzed.

## 2. Experimental details

All coatings were deposited at 550 °C in a lab-scale deposition system (diameter 0.44 m and height 0.36 m, plus flanges) evacuated to a base pressure of ~$2.7 \cdot 10^{-6}$ Pa ($2 \cdot 10^{-8}$ Torr). The system is equipped with four, two inch-in-diameter, unbalanced and magnetically coupled magnetron sources directed towards the substrate from below, with a 30° angle from the substrate normal and a source-to-substrate distance of 0.14 m. The substrate rotation speed was set to 30 rpm for all depositions. $(Al_{1-x}V_x)_2O_3$, $0 \leq x \leq 1$ coatings were co-sputtered using two opposing magnetron sources, one equipped with a ¼ inch (6.35 mm) thick Al target (99.99% Kurt J. Lesker) and the other with a ⅛ inch (3.175 mm) thick V target (99.5% Kurt J. Lesker). Each magnetron was powered by an Advanced Energy (AE) DC MDX 500 W power supplies in series with (AE) Sparc-le V pulsing unit. The MDX supply was run in power control mode. The Sparc-le V was in self run mode (continuous voltage reversal according to pulsing frequency) at 50 kHz pulsing frequency and 2 μs positive pulse time with 10% of the negative voltage as positive reverse voltage magnitude (RVM). The resulting duty cycle (negative pulse) was 90 %. The bias voltage was controlled by a ES 0300-0.45 power supply from Power box connected in series with a Sparc-le V pulsing unit operated in a self-run mode at 100 kHz frequency and 1 μs reverse time with 10% positive RVM. -120 V or -20 V was applied as the negative bias value during the pulsed DC bias. The bias pulse has been verified with the help of an oscilloscope measuring at a T-connection mounted on the chambers bias feed through, the given bias values being the set values as well as the measured average values, related to chamber ground. The bias was applied through a Mo sample holder that was sand blasted between each run. 18×18 mm pieces of a p-doped (Boron, 10-20 Ωcm) Si(100) wafer was used as substrate and placed in the center of the Mo sample holder. Prior to loading, the substrates were ultrasonically cleaned in acetone and ethanol, 10 min each. The total argon flow was set to 36.4 sccm and divided on the two active targets in all depositions, resulting in 0.27 Pa (2 mTorr) Ar partial pressure. The oxygen was distributed from a gas outlet on the chamber wall at the height of the substrate. The pure $Al_2O_3$ and $V_2O_3$ coatings (when only one source was ignited), were optimized to be deposited in the transition zone after hysteresis studies varying the oxygen flow. The oxygen flow for Al and V was chosen so that the target voltage was -14% and 16 % from the metal values, respectively (measured on the full voltage change between metal-poison mode). Due to the high effective pumping speed 250 l/s (calculations based on [39]) the hysteresis effect was small and the depositions conditions were stable even in the transition region. The intermediate compositions, with both magnetrons running were controlled by keeping the $O_2$ partial pressure constant at 0.01 Pa (0.08 mTorr) through manual slight adjustment of the $O_2$-flow during the deposition. The general deposition procedure was that each deposition started with sputter cleaning of the target in Ar to obtain a stable voltage. Oxygen was then introduced, followed by an additional delay to reach stable voltage and pressure (the shutters in front of the active targets were open during the entire process). Finally, the sample shutter was opened and the deposition started. The main sample series was a constant -120 V bias series with varying Al/V ratio by changing the power to the respective magnetron. For three of these compositions -20 V bias coatings was also done in order to study the effect of the bias. In the paper the samples will be addressed with the measured V at.% metal fraction, bias level and sample letter. The different process conditions are summarized in table I.

Top-view and fracture cross-sectional view micrographs of the samples were obtained with a scanning electron microscope (SEM), Leo 1550 Gemini field emission gun operated with 5 kV acceleration voltage and an in-lens detector.

The chemical composition was mainly characterized by elastic recoil detection analysis (ERDA) using 36 MeV $^{127}I^{8+}$ primary ions provided by the tandem accelerator at Uppsala University, Sweden. The detector was a combined TOF-E spectrometer equipped with a solid-state detector. The recoil detection angle between incident beam and detector telescope was 45º with an incidence angle of 67° with respect to the sample surface normal. The information depth was typically around ~300 nm. The achieved precision of the measurement for oxygen was ~1% (mainly due to experimental statistics) and the detection limit for recoil species lighter than ~100 a.m.u. was ~0.1%. The data were evaluated with POTKU version 1.1. Further description of the technique, experimental facilities and the achievable accuracy can be found elsewhere [40-42].

In addition to ERDA, Rutherford backscattering spectrometry (RBS) was performed on selected samples to improve the overall accuracy of the data. 2 MeV $^4He^+$ primary ions were employed as a projectile. A solid state detector positioned at a scattering angle θ of 170° was employed to record energy spectra. For the bulk sapphire reference, small arbitrary rotations of the sample by ± 2° around the equilibrium position were employed to reduce possible channeling effects. Spectra were evaluated with SIMNRA 7.01 with a statistical error of ~1 %. Due to the known uncertainty in electronic stopping powers for the early transition metals SIMNRA, analysis of the V-rich samples was performed using updated values for He in V, equivalent to a 5% change in stopping power of V compared to tabulated values in the program [43].

Phase identification and crystal structure of the as-deposited coatings were characterized by XRD in a Panalytical Empyrean equipment, using Cu $K_\alpha$ irradiation. Both θ-2θ and grazing incidence x-ray diffraction (GIXRD) (ω=1.5°) scans were performed, using a parallel beam mirror with ¼° divergence slit, a fixed mask of 4 mm on the incident x-ray beam side, a 0.27° parallel plate collimator, 0.02 mm nickel foil and a PIXcel$^{3D}$ detector in open detector mode on the diffracted x-ray beam side. A detector step size of 0.04°, and count time 5.28 sec was used for all measurements in this setup. The unit cell of selected samples was calculated using the "UnitCell" software [44]. The 2θ angles input comes from $K_{\alpha 1}$ peak refinements data done with XRD-program "Highscore" from Malvern Panalytical.

The annealing experiments were performed in air using an Anton Paar DHS 1100 heating stage (equipped with an AlN heating element) mounted on a Panalytical Empyrean XRD equipment. The heating stage could go up to 1100 °C and was used with a heating/cooling rate of 20 °C/ min with a 1 h isothermal hold time at the annealing temperature. For the samples annealed at 1000 °C a 2h and 50 min hold time was experienced due to a malfunctioning in the temperature control program. The samples were 3×5 mm in size and could all be placed next to each other, on the heating stage under a carbon dome. However, only one sample at a time could be characterized by *in situ* XRD. The same samples were used for all cycles until they had oxidized to such extent that no further mechanical characterization was possible.

Light optical microscopy (LOM) was used to photograph the sample between each annealing run. Polishing of the film surface has caused interference fringes visible in LOM photos due to variable coating thickness at the boarder of the samples.

For the smaller samples used in the annealing study, only grazing incidence XRD was used with an incident angle of 1.5°, but with a different optical setup in order to accelerate the measurement. To illuminate only the small sample and not the heater plate, the incident beam side was equipped with parallel beam mirror with 1/32° divergence slit and 4 mm mask. On the diffracted beam side, a 5 mm anti-scatter slit was positioned close to the carbon dome (11.5 mm from detector), followed by a 0.02 mm nickel foil and the same PIXel$^{3D}$ detector, run in scanning line mode (255 number of channels covering a 3.347° 2θ range). A detector step size of 0.0525° was used. The second setup made it possible to perform a full scan, 2θ 15-90 °, in 45 min on such a small sample with sufficient counts. The diffuse scatter of the primary beam, however, resulted in an increased background for smaller 2θ angles, <25°.

Nanoindentation hardness measurements were performed in a Hysitron, TriboIndenter 950 equipped with a 2D-transducer using a diamond Berkovich tip with a nominal radius of 100 nm. The tip area function (AF) was calibrated by indentation with variable load (100-11000 μN) in fused quartz prior and after measurement of the samples for each annealing temperature. This resulted in indentation depth of ~19-200 nm with a valid AF calibration for ~40-200 nm indentation depth. The stated accuracy for the equipment is 10 % for hardness and 5 % for Young's modulus for the fused quartz reference. Prior to the first nanohardness evaluation, before the annealing study, the samples, cut to 3×5 mm size, were gently polished with a 0.1 μm I diamond LF 3 mil disc (Allied, High Tech Products). This procedure caused a tapered surface at the rim of the cut sample piece where the Si substrate became visible. The center of the sample was not significantly polished. The hardness was measured close to maximum coating thickness but in the polished region. The load was chosen such to obtain indentation depth below 10% of the total coating thickness, ~70-90 nm depth. A test with a constant 4 mN load over the tapered surface all the way to the Si substrate confirmed that the measured values were not affected by the substrate until the tip being laterally very close to the exposed Si-substrate. The used nanoindentation method was quasi static, open loop, with a 8 sec load and unloading ramp giving a loading rate of 500 μN/sec with a 5 sec hold at max load. The measurement equipment has an automatic correction for thermal drift prior to each indent. The hardness was evaluated according to the Oliver and Pharr method [45], using upper 95 to lower 20 % of the unloading segment for the unloading stiffness fit procedure (for Si substrate 95 to 30% due to phase transformation when unloading [46]). The polished Si on the rim of the pure $Al_2O_3$-sample is the internal reference of the hardness measurement, making it more reliable to evaluate the hardness and modulus over extended time as is demanded for such an annealing study. 50 measurements (2×25 along the interference fringes, constant thickness lines) were done at the same load on each sample (30 indents for $SiO_2$ and Si). Out of these normally ~47 was used for evaluating the stated average values if the coating had not started to oxidize (all were used for $SiO_2$ and ~27 for Si). When the hardness values started to decrease due to further oxidation and the standard deviation increases, the number of used indents are fewer.

## 3. Results
### 3.1. Synthesis and characterization of as-deposited coatings
#### 3.1.1. Chemical composition

Table II lists the chemical composition of the as-deposited (Al,V)O coatings measured with ERDA (white rows) and RBS (grey rows) as well as the coating thickness measured in fractured cross-section SEM micrographs. The oxygen content by ERDA and RBS as a function of V/(V+Al) % for all measured samples is shown in Figure 1. The ERDA measurements show a slightly higher oxygen content for the Al-rich oxide compared to the V-rich oxide coatings, especially the pure vanadium oxide. However, compared to a stoichiometric metal oxide compound with a $Me_2O_3$ composition, marked with dotted line in Figure 1, all coatings are measured sub-stoichiometric with respect to O. As apparent the ERDA measurement of the bulk (001) sapphire substrate underestimates the oxygen content by approximately 2.5 at.%. Similar, the deviation in oxygen content for the pure Al-oxide coating (sample H), with respect to the sapphire substrate, is about 0.7 at.% as measured with ERDA. This is within the statistical uncertainties of the depth averaging performed for each measurements. For the vanadium oxide coating (sample A), however, the deviation in O content is larger, about 1.9 at.%, and is therefore interpreted as a sub-stoichiometric $V_2O_3$ oxide. To further verify the film composition, RBS measurements were made on selected samples. RBS indicated stoichiometric O-content for the bulk sapphire sample and 1-2 at.% sub-stoichiometry for the Al-rich oxides, i.e., within the error margins. The vanadium oxide sample, on the other hand, remains significantly sub-stoichiometric at 55.7 at.% oxygen instead of the stoichiometric 60 at.%.

The argon content (Table II) is about 1 at.% for the Al-rich coatings deposited with a -120 V of bias (samples A to H) but is reduced to ~0.5 at.% with increasing V-content. For the -20 V bias coatings (samples I to K) the content is even less, close to the detection limit for ERDA (0.1 at.%). The carbon contamination is close to the detection limit for ERDA and not detected by RBS, which indicates clean deposition conditions.

### 3.1.2. Coating thickness and microstructure

The (Al,V)O coating thicknesses as shown in Table II are about 1 µm in average with a maximum deviation of 170 nm for -20 V 180/70 coating (sample J). The deposition rate varied significantly between the different power ratios used for obtaining the gradual shift in V/(Al+V) composition. While the pure $V_2O_3$ and $Al_2O_3$ oxides showed the lowest deposition rates of 466 and 573 nm/h, respectively, the highest deposition rate of 1210 nm/h was observed for the composition using the maximum total power of 180/100 (sample F). To compensate for the different rates, the deposition time for each sample was adapted, in order to obtain coatings of similar thickness.

Figure 2 shows (top row) surface morphology and (bottom row) cross sectional SEM micrographs, obtained with the in-lens detector and an acceleration voltage of 5 kV, of (Al,V)O coatings deposited with a substrate bias of -120 V and decreasing metal fraction V/(Al+V) from (left) 100 at.% of V (sample A), i.e., $V_2O_3$, to (right) 0 at.% of V, i.e., $Al_2O_3$ (sample H). The coatings may be divided into four characteristic groups having a columnar structure and gradually decreasing grain size with decreasing V-content: Pure $V_2O_3$, with large grains, the coatings containing 63 and 52 at.% of V metal fraction (samples B and C) with elongated grains, the coatings containing 47, 42 and 18 at.% of V (samples D, E and F) with equiaxed grains and the coatings containing 7 and 0 at.% of V (samples G and H) having a cauliflower-like structure comprising of smaller grains. The latter structure, i.e., for $Al_2O_3$ generates an increased coating roughness.

Figure 3 shows surface morphology (top row) and cross sectional SEM micrographs (bottom row), obtained with the in-lens detector and an acceleration voltage of 5 kV, of (Al,V)O coatings deposited with a substrate bias of -20 V and decreasing V metal fraction from (left) 63 at.% of V (sample I) to (right) 0 at.% of V (sample K). The coatings prepared with a lower substrate bias show a similar columnar structure as for the -120 V coatings with gradually decreasing grain size with reduced V-content. The -20 V coatings show however a slight increase in porosity compared to the -120 V coatings. The small addition of 7 at.% of V (sample G and J) decreases the roughness significantly and results in densification of the microstructure, both for -20 V and -120 V bias.

### 3.1.3. Crystal structure from XRD

Figure 4 shows GIXRD (1.5°) of a) the -120 V bias series and b) the -20 V bias series with increasing V-content from bottom to top. For the -120 V bias series, the diffractogram acquired on the pure vanadium oxide coating (sample A) exhibits diffraction peaks matching the corundum structured $V_2O_3$ phase in accordance with the Powder Diffraction File (PDF) 00-034-0187, marked with fine dashed lines. Comparing the unit cell of the PDF-file with what is obtained from fitting the 20 visible peak positions gives a unit cell which correspond to the second digit with the pdf-card: a =4.956 Å, b=4.955Å, c=13.917 Å. This results in a unit cell volume that is marginally smaller (0.6%) than specified in the PDF-card.

The pure $Al_2O_3$ (sample H) reveals a good match to the cubic defect spinel structure, $\gamma$-$Al_2O_3$, space group Fd-3m (PDF-card 00-010-0425. Note that there are two PDF cards, where the more recent 00-050-0741 contains fewer planes and differ by 0.2-0.6% in 2θ angle). The notation pure refers to the chemical composition rather than the phase purity. The well-defined diffraction peaks at 39.47° attributed to $\gamma$-$Al_2O_3$ 222 and higher order diffractions (Figure 4 a) are situated in between the values stated in these two pdf-cards. However, the 111, 220 and 311 peaks at lower angles deviate more, 1.2-1.7 %, which may be related to reduced long range order. This is also translated into low and broad intensity, especially for the 111 and 220 peak. Nevertheless, the observation of all these peaks is a strong indication of an obtained $\gamma$-$Al_2O_3$ structure. Considering that all peaks are visible, the crystallinity of the sample is high, still the presence of some amorphous fraction cannot be excluded.

No peak shifts are observed between θ-2θ scan (not shown) and the GIXRD scan for the well-defined 222, 400 and 440 peaks. This indicates that there are no large macroscopic stress effects in the coating. The strongest diffraction peaks, 400 at 45.774° and 440 at 66.819°, are from where the oxygen cubic close-packed sublattice and octahedral coordinated Al 400, octahedral and tetrahedral Al 440 scatter in phase, respectively [47]. These two peaks remain the most prominent peaks as V is incorporated into the coating. Upon adding V, the unit cell expands moving the diffraction peaks to smaller angles. The long-range order (low angle peaks, 222 and below) is quickly lost when adding V, see Figure 4, diffractogram 7 and 18 at.% of V (samples G and F respectively).

When performing the same unit cell fitting for the Al-rich side (as for $V_2O_3$), with increasing V-content, there is a gradual shift to larger unit cell, indicating a solid solution between Al and V. Assuming a tetragonal unit cell. This results in an expansion of a by 0.9 % and an unit cell volume expansion of 2.9% by adding 18 at.% of V (sample F), see Table S1 in supplementary material.

For the coatings deposited at -20 V bias with low or no V (samples J and K), Figure 4 b), the 400 and 440 peaks shift significantly to larger angles compared to corresponding coatings deposited at -120 V bias, aligning better with the PDF-card value. Especially for the 9 at.% of V (sample J), there is not as strong shift to lower angles as for the corresponding -120 V bias coating. Also in this case, the effect of macroscopic stresses is limited since there is no shift in the 400 peak position between GIXRD and θ-2θ scan (not shown) for the pure $Al_2O_3$-coating. The overlayered XRD diffractogram of the -120 V and -20 V coatings, showing clear peaks shifts, can be found in supplementary material, Figure S1.

Further addition of the V (42 at.% of V, sample E) for the -120 V biased coatings results in a broad 400 peak and not as clear shift to smaller angles with further increasing V. The coatings with 42, 47, and 52 at.% of V (samples E, D and C) were deposited in order to evaluate the possibility to synthesize the previously reported AlVO$_3$ structure [32]. The positions of the AlVO$_3$ related diffraction peaks reported in the pdf-card 00-025-0027, are marked with long dashed line in Figure 4 a). The sample with 63 at.% of V metal fraction (sample B) exhibits several peaks 33.627°, 36.522°, 41.472°, and 50.4° matching the α-V$_2$O$_3$ 104, 110, 113 and 024 peaks respectively, but at slightly higher angles compared to the PDF-card. In addition to this, a broad peak is visible in the region of the γ-Al$_2$O$_3$ 400 peak position at ~46.2°. This peak cannot be explained with only the α-V$_2$O$_3$ structure. For the corresponding -20 V bias coating with 63 at.% of V metal fraction (sample B), the same diffraction peaks are observed, albeit significantly sharper and with higher intensity. Their positions are slightly shifted to larger angles compared to the -120 V case.

Overall, the XRD of the as deposited -120 V coatings show well-developed γ-Al$_2$O$_3$ and α-V$_2$O$_3$ for the pure phases. Adding V shifts the γ-Al$_2$O$_3$ main peaks to lower angles as would be expected by increased lattice spacing due to V substitution. Clear α-V$_2$O$_3$ peaks, shifted to larger angles, indicating an Al-solid solution, are visible for the 63 at.% of V metal fraction coating (sample B).

### 3.2. Annealing in air of coatings with high Al content
#### 3.2.1. Light optical microscope images of annealed coatings

Coatings with a high Al content, having the highest as-deposited nanoindentation hardness (see section 3.2.3), were selected for the annealing experiments. The same sample for each coating was cycled to increasingly higher temperature. After each annealing step, LOM, XRD and nanoindentation characterizations were performed. Figure 5 shows LOM images of the coatings after annealing to different temperatures. Coatings with higher V contents are shown from top to bottom and a higher annealing temperature from left to the right. The photographed area for each sample and temperature is in the corner part of the 3×5 mm cut samples where the polished region contacts the Si-substrate (left in each image). The gradual reduction in coating thickness at the border of the samples, due to polishing of the as-deposited coatings, causes the interference fringes visible on all coatings prior to severe oxidation. The images show that the coating with 42 at.% of V (sample E) has started to oxidize severely already slightly below the growth temperature at 500 °C. The hardness was, however, still possible to evaluate and therefore it was annealed also at 600 °C resulting in larger 3D oxide structures (yellow). As an extra test to see if it was possible to form $AlVO_3$ upon annealing, the coating with 52 at.% V metal fraction (sample C) was added for a one step annealing run to 700 °C. This resulted in large 3D oxide formation, seen on the top row (right). The coating containing 18 at.% of V (sample F) remained unchanged up to 600 °C. After annealing to 700 °C, however, elongated oxide features formed (right). After annealing to 800 °C the oxide started to exhibit dark 3D-structures (right), while the base oxide parts of the coating remained seemingly unaffected with intact thickness fringes. The coating with 7 at.% of V (sample F) remained unchanged up to 800 °C, but after annealing to 900 °C, an evenly distributed pattern of round features covered the surface. Each round region, possibly showing a stress field, has cracks extending from the middle of each round region (reasons further treated in the discussion, section 4.4). The pure $Al_2O_3$ coating (sample H), bottom row, does not show any change in the coating structure up to 1000 °C but the visible Si-substrate has experienced a small color change indicating initial oxidation. After annealing to 1100 °C, most of the coating has spalled (right-hand side of the image) except for in the edge region (middle of the image). The exposed Si-substrate is now bluish due to thermal oxidation of Si to $SiO_2$ (left).

### 3.2.2. Crystal structure development with annealing temperature

The corresponding XRD-patterns from the annealed samples are shown in Figures 6 (a-d). Each figure shows the diffraction patterns for one composition with increasing annealing temperature. The higher background for the sample up to ~30° is due to the XRD-setup and diffuse scattering of the primary beam at low 2θ-angles.

Figure 6 a) shows the XRD pattern for the coating with 42 at.% of V (sample E) which starts to oxidize at 500 °C. The peaks match the orthorhombic $V_2O_5$ phase (PDF-card 00-041-1426, all significant peaks are marked with double dashed lines). Some additional peaks appear more strongly after annealing to 600 °C, i.e., the $V_2O_5$ 200 peak at ~15.4°. In conjunction with the $V_2O_5$ phase formation on the surface of the coating, visible in the optical images in Figure 5 (correlated V-enrichment confirmed with top view SEM-EDX, not shown), the initial γ-$(Al,V)_2O_3$ structure reverts to match the 400 and 440 peak positions of the pure γ-$Al_2O_3$ phase. There are also 3 new diffraction peaks at around 28°, after annealing to 600 °C, which is not possible to match to $V_2O_5$ or γ-$Al_2O_3$, but rather match the main peaks for the $AlVO_4$ phase (PDF-card 00-039-0276). XRD acquired on the coating with 52 at.% of V (sample C) directly annealed at 700 °C, exhibiting large 3D-formation (LOM displayed in figure 5, top right), shows evidence of predominant formation of $AlVO_4$ (diffractogram not shown).

Figure 6 b) shows the XRD pattern for the coating with 18 at.% of V (sample F). The XRD remains unchanged until visible oxidation occurs after annealing to 700 °C. The new main diffraction peaks match the $V_2O_5$ structure, like what was observed for the coating with 42 at.% of V (sample E), but with fewer diffraction peaks. This could be explained by a less developed phase formation or texture. The $V_2O_5$ phase develops after annealing to 800 °C. The γ-$Al_2O_3$ structure shifts to higher angles after annealing to 700 °C and is positioned at the γ-$Al_2O_3$ pdf-card values after annealing to 800 °C, now including four diffraction peaks compared to two peaks observed for the as-deposited coating with 42 at.% of V (sample E).

Figure 6 c) shows the XRD pattern for the coating with 7 at.% of V (sample G) which remains in a solid solution γ-(Al,V)$_2$O$_3$ structure, shifted to lower angles than pure γ-Al$_2$O$_3$, up to annealing at 700-800 °C. After annealing to 800 °C the 311 diffraction peak at 37.6° is significantly more visible and after annealing to 900 °C, the peaks matches γ-Al$_2$O$_3$ PDF-card (values marked in blue). In addition to the γ-Al$_2$O$_3$ phase a small peak matching the V$_2$O$_5$ 400 at 31.2 °C is observed, however the 200 and 600 peaks are not visible. This phase forms in conjunction with the loss of V from the γ-Al$_2$O$_3$-structure, at lower annealing temperatures. e.g. after annealing to 700 °C there is an unusual high intensity peak at the position matching the 600 peak for V$_2$O$_5$ at 47.38°, which disappear for higher annealing temperatures.

Figure 6 d) shows the XRD pattern for the pure γ-Al$_2$O$_3$ coating (sample H), in good agreement with the diffraction peaks in the pdf-card 00-010-0425 (blue lines). This coating essentially remains unchanged up to the highest annealing temperature of 1100 °C, where small amounts of α-Al$_2$O$_3$ can be detected (pdf-card 00-046-1212 marked with grey dashed lines). Due to the flaking of the transformed coating, the diffracted intensity for the corundum phase is drastically reduced. Below the maximum annealing temperature, the γ-Al$_2$O$_3$ crystal quality seems to improve with increasing annealing temperature. The change is largest in the temperature span 800-1000°C. This is best exemplified by following the FWHM of the 311 diffraction peak at 37.6°. Starting at 2.6° on as-deposited coating, the FWHM of the 311 diffraction peak decreases to 2.4° after annealing to 800°C and to 1.6° after further annealing to 1000°C.

Based on the above, the V-rich phases, i.e. coatings with 42 and 18 at.% of V metal fraction (samples E and F), started forming V$_2$O$_5$ after annealing to 500 °C and 700 °C, respectively. In conjunction with this, their primary γ-Al$_2$O$_3$ 400 and 440 diffraction peaks are shifted back to pure γ-Al$_2$O$_3$. The coating with 7 at.% of V (sample G) started to revert to γ-Al$_2$O$_3$ after annealing to 800 °C without clear indication of any type of vanadium oxide forming. The pure γ-Al$_2$O$_3$ (sample H) revealed the highest temperature stability of the coatings in this study. α-Al$_2$O$_3$ is formed in conjunction with coating flaking after annealing at 1100 °C.

### 3.2.3. The evolution of mechanical properties with annealing temperature

Figure 7 shows the nanohardness a) and reduced elastic modulus b) of all coatings, evaluated from nanoindentation measurements after each air annealing step. The coating with 42 at.% of V (sample E exhibits a dramatic drop in hardness to 12.5±2.7 GPa after annealing at 500 °C. At the same time, the modulus remains close to its initial value but with a significant larger spread. The hardness values for coatings showing new surface oxides in Figure 5, should just be taken as an indication of the coating hardness, primarily originating from the base coating hardness, showing the thickness interference fringes. This is based on the assumption that most invalid curves, which are not included in the average values, originates from indentations on protruding surface oxide crystals. After annealing at 600 °C, it was not possible to evaluate the mechanical properties further due to the significant 3D oxide formation on the surface.

The coating with 18 at.% of V (sample F) retains its hardness of 22±0.6 GPa and reduced modulus of 225±4.1 GPa until after annealing at 600 °C. At higher temperature, the onset of V$_2$O$_5$ formation after annealing at 700 °C, resulted in a dramatic decrease in coating hardness while the reduced modulus only show a small decrease. The hardness and reduced modulus further decreased to 14.8 ± 2.1 GPa and 207±31 GPa, respectively, after annealing at 800 °C.

The highest hardness of 23.5±1 GPa was obtained for the coating with 7 at.% of V (sample G) and the pure γ-$Al_2O_3$ coating (sample H) and which remained high until the coatings exhibit grain growth and / or further oxidization. The hardness of coating with 7 at.% of V (sample G) starts to decrease already after annealing at 800 °C, that is, before any visible oxidation is observed in LOM (see Figure 5). The corresponding XRD show a small shift to larger angles. The spread of the hardness is still small but the loss in hardness, although small, is significant. It is not until after annealing to 900 °C, accompanied by visible surface morphology change in LOM, that the hardness drops to 16.1 ± 2.3 GPa in conjunction with increased spread. Top view SEM show some larger crystals on the surface of the coating (not shown). The reduced elastic modulus follows the same pattern but the decrease is significantly smaller.

The pure γ-$Al_2O_3$ sample shows a hardness reduction after annealing to 1000 °C and the coating spalled after annealing to 1100 °C, preventing hardness measurement after annealing to that temperature.

## 4. Discussion

### 4.1. Chemical composition

The chemical composition presented in Table II and Figure 1 indicates a varying degree of sub-stoichiometry for the deposited coatings, which increases with increasing V-content. ERDA, while being much more sensitive for light species, is more prone to systematic uncertainties originating from two sources: at first, for light species, the detection efficiency deviates significantly from unity; secondly, uncertainties in the electronic stopping powers of the primary and recoil species are entering the evaluation. While ToF-ERDA and RBS are both ion beam methods, featuring similarly high accuracy in cross sections, the above mentioned factors do not affect the accuracy of RBS. Note, in this context, that a recent study determining electronic stopping powers for $H^+$ and $He^+$ ions in V could show that tabulated values were differing by up to 10 % from the results of the new measurements [43]. As in particular stopping powers for heavier ions as well as their predicted values in compounds are often based on extrapolation from known data for light ions and monoelemental targets, the achievable accuracy in ERDA for compounds containing V can thus be considered to be limited.

From this background, the RBS data is judged to be more accurate with respect to the metal-oxygen ratio than the ERDA measurements with RBS showing that the bulk sapphire has a stoichiometric composition of Al/O of ~0.67. The $Al_2O_3$-rich coatings measured with RBS show a 1-2 at.% deficiency of oxygen compared to the bulk sapphire. The additional discrepancy in oxygen content between RBS-values and ERDA of on average ~1.5 at.% is thus considered a systematical shift between these techniques, which lead then to the interpretation that all Al containing coatings (samples B to H, I to K) are slightly (1-2 at.%) sub-stoichiometric. The exception from this trend is pure $V_2O_3$, which shows a larger decrease in oxygen content in both RBS and in ERDA compared to the Al-rich alloys. This coating shows comparable coarse grained structure compared to the rest of the coatings, Figure 2, with a XRD-pattern well aligned with pure α-$V_2O_3$, se Figure 4 a).

The Ar content is ~1 at.% when bias is set to -120 V. The lower Ar-content for the $V_2O_3$ – rich coating may indicate higher diffusivity and Ar non-sticking during growth. The larger crystals seen for these coatings (Figure 2), compared to the Al-rich oxides, support this reasoning.

### 4.2. Micro structure

The gradual crystal size reduction seen with SEM, Figure 2, when increasing the Al-content in the coating is not possible to explain only based on the small difference in melting point between the binary oxide: pure $V_2O_3$ melts at 2249 °K and pure $Al_2O_3$ melts at 2327 °K [48]. This gives a slightly larger degree of undercooling for Al-rich pseudo-binary oxide with more nuclei forming in combination with lower diffusivity, possibly explaining the observed change in microstructure. Inherent differences in diffusivity between these oxides may be a more plausible explanation for the observed difference in microstructure. The diffusion mechanism observed in α-$Al_2O_3$, thoroughly reviewed by Heuer [49], shows a large variation in measured diffusion coefficient for different types of diffusion and alloying. Other work treating the diffusion and formation of protective $Al_2O_3$, $Cr_2O_3$ and $SiO_2$ oxide scale was performed by Stott et al. [3, 50] .

The microstructure changes significantly between the coatings deposited with -120 V and -20 V bias, with increased porosity for the -20 V bias case. For pure $Al_2O_3$ with -20 V bias the increased diameter of the protrusions with increasing coating thickness, in combination with lower diffusivity due to lower bias (lower energy bombardment), lead to shadowing which causes porosity around the larger features, as seen when comparing Figure 3 to Figure 2.

### 4.3. XRD on as deposited films

The crystalline structure of the as-deposited $Al_2O_3$ coatings (sample H and K) was attributed to the γ-phase. However, due to similarities between various $Al_2O_3$ polymorphs, phase identification is difficult. Zhou [47] suggested that the γ- or η-phase in a powder sample, based on their difference in x-ray form factor, could be identified by comparing the 311 and 222 diffraction peak intensities at 37.21° and 39.48°, respectively. Texture effects, potentially disturbing this interpretation is ruled out by a symmetric θ-2θ scan on the same sample (not shown) which show similar intensity distribution as observed in Figure 4. On that basis, the reasoning from Zhou is judged to be valid also for this coating, and the structure is concluded to have a gamma structure.

In the perfect γ-$Al_2O_3$ structure, the fully occupied and evenly spaced oxygen layers are arranged perpendicular to the [111] direction. In between these oxygen layers, every second layer contains octahedrally coordinated Al-atoms and every second layer contains a mixture of tetragonal and octahedrally coordinated Al-atoms. The difference in diffraction peak broadening, even for lattice planes from the same crystal zone, like 111, 222, 333 and 444, is typical for these kind of structures according to Zhou [47] and has its origin in alternating layers along the [111]-direction. The distinct layering in the pure gamma phase explain why the 111 and 222 are lost first while the 444 remains even after having introduced about 18 at.% of V in the structure (sample F - shown Figure 4 a). At this composition, the periodicity is reduced to the shortest possible distance; that between close packed oxygen and tetrahedrally coordinated Al.

The increased broadening and decreased intensity of the 220 diffraction peak upon adding V show the increased short range disorder in the structure. This is because the 220 diffraction peak from the γ-$Al_2O_3$ only originates from tetrahedral coordinated interstitials on every second metal layer, which according to Reid [32] only is accessible for $Al^{3+}$ due to the larger size of $V^{3+}$. If tetrahedral holes would start to be occupied by $Al^{3+}$ to some degree on every metal layer (implying octahedral vacancies in those layers as well) the 220 periodicity would disappear but the 440 would remain. This is also what is observed when adding V up to 18 at.% .

The fact that the main diffraction peaks for the gamma structure shift towards the peak positions of pure γ-$Al_2O_3$ for the coatings deposited with a -20 V bias indicates that the more energetic bombardment, promoting atomic mixing when using -120 V bias, is needed in order to incorporate

V in the γ-Al$_2$O$_3$ structure. This is understandable from the perspective that both metals prefer to be in octahedral interstitial position and additional mixing is needed in order to promote more Al into tetrahedral positions [32].

Adding more V than 18 at.% metal fraction (sample F) change the diffractogram to seemingly lose the symmetry in the [111] direction, as 444 becomes week, broad or non-existing. As mentioned in the results section, the range of equal fraction of V and Al aimed to investigate if the previously reported AlVO$_3$-phase would form [32].

The apparent match to AlVO$_3$ of the 222 and 531 diffraction peaks at ~36.7° and ~65.4° respectively (in Figure 4 a) come from planes which should have low intensities accordingly to the pdf-card (00-025-0027) and are therefore unlikely to be related to this phase. Instead, it seems likely there is a gradual transformation from the gamma defect spinel structure to the α-V$_2$O$_3$ corundum structure with increasing V content to around 63 at.% V metal fraction. The 440 gamma peak approaches the position of the 300 corundum peak with increasing V-content. The 400 has shifted to lower angles until 52 at.% of V (sample C), then shift to larger diffraction angles with increasing V-content, especially visible for the coating with 63 at.% V (sample B) at 46.11° comparable to 45.863° for a pure γ-Al$_2$O$_3$ (sample H). There is also a small peak at 67.3°, to the right of the 300 corundum peak in sample H with 63 at.% V, which match the 440 peak shifted to slightly larger angles than pure γ-Al$_2$O$_3$. The peaks at 33.627°, 36.522°, 41.472°, 50.4° matches the α-V$_2$O$_3$ 104, 110, 113 and 024 planes respectively but at slightly higher angles, possibly due to some degree of solid solution with the smaller Al-atom. The 012 diffraction peak at 23.6° is shifted to lower angles, this is surprising since the 024 at 50.4° is shifted to higher angles compared to the pdf-card position of α-V$_2$O$_3$. The solution to this behavior may be found in the corresponding -20 V bias diffractogram discussed below.

The diffractogram acquired on the corresponding -20 V bias coating with 63 at.% of V (sample I) exhibits significantly more of sharp and high intensity peaks. Peak splits are observed for the 012 and the 024 peaks, se Figure 4 b). The intensity for the two sets of peak split are reversed between the two set of planes. Calculating the difference in lattice parameter for the two sets of peak splits gives a 3.95% and 4.19% plane spacing difference respectively. This indicates that they indeed come from the same set of planes. The reverse magnitude in intensity between the two set of planes is attributed to ordering at the metal sublattice. There are also two sharp high intensity peaks, 400 and 440, attributed to the γ-like $Al_2O_3$ structure but shifted to slightly higher angles compared to the γ-$Al_2O_3$ pdf-card, as well as compared to the corresponding peaks from the -120 V bias coating. Calculating the a-lattice parameter for these two planes, adding also the 400 measured from a θ-2θ scan (not shown), results in a=7.827 Å with a ± 0.08% standard deviation which is 0.9% smaller than what is stated in the pdf-card. These peaks cannot be explained with the α-$V_2O_3$-structure, thus a two-phase structure of α-$(V,Al)_2O_3$ and γ-like $Al_2O_3$ is likely to be present in this sample. The high number of peaks above 70° are too many to properly match any of the two proposed structures without detailed description of the possible ordering of the metal sublattice which is left for further studies on the topic. Better understanding of the often complicated crystal structures obtained in the as-deposited coatings could possibly be obtained by means of high resolution energy filtered cryo TEM as previous shown for γ-$Al_2O_3$ coatings by Engelhart et al. [51].

### 4.4. Air-annealed coatings

The annealed coatings show further oxidation taking place in all the V-alloyed samples, at increasingly higher annealing temperature with lower as deposited V-content. Previous reports on the oxidation onset of vanadium-based nitrides show that $V_2O_5$ forms at ~500 °C and that its melting point is ~680-700 °C [27, 37]. The $Al_2O_3$-$V_2O_5$ pseudo binary phase diagram [31] show melting of the pure $V_2O_5$ at ~670 °C but with possible lower melting point, 640 °C, when alloyed with $Al_2O_3$ up to 50 mol %. $AlVO_4$ is said to melt and decompose into $Al_2O_3$ and liquid V-rich phase at 745 °C but there are several different temperatures reported, ranging from 625 to 778 °C, depending on synthesis rout and annealing conditions [31]. In the present work, $AlVO_4$ was obtained (LOM in Figure 5, XRD not shown) after annealing 52 at.% metal fraction V coating (sample C) for 1 h at 700 C. This indicates that given the present deposition and annealing conditions, the higher decomposition temperatures are the correct one for the here studied coating. Reports from oxidation studies with tracer $O^{18}$ in AlCrVN show that oxygen is diffusing through the formed $AlVO_4$ and $V_2O_5$ oxides [52] which could explain the observed 3D grown oxides of the 52 at.% metal fraction V coating in this study.

Thermal oxidation in air at 300-500°C of a $V_2O_3$ film gradually transformed into $V_2O_5$ starting at 400 °C and finally converted at 500 °C [53]. This result, in combination with previous reports on onset of $V_2O_5$ oxidation in the TiAlN:V-system at around 500 °C, correlates well with formation of $V_2O_5$ after annealing the 42 at.% V metal fraction coating (sample E) at 500 °C. Further annealing of the coating at 600 °C yield well developed $V_2O_5$ and fully reverted γ-$Al_2O_3$ main diffraction 400 and 440 peaks. The oxide coverage is however not homogenous according to Figure 5, but rather formed in elongated crystals, especially visible after annealing at 500 °C.

The onset temperature of oxidation can be pushed to higher temperatures, 700 °C seen in Figure 6 b), by lowering V-content further to 18 at.% V metal fraction (sample F). Then the onset temperature of $V_2O_5$ formation is higher than for i.e. 25 at.% V metal fraction containing TiAlVN coating (~600-650 °C) [37]. The formed oxide, visible as silvery and brown regions in Figure 5, cluster inhomogenously in regions of closely spaced crystals, protruding from the as grown surface, visible in LOM and SEM (not shown). The high intensity $V_2O_5$ diffraction peaks are absent during in-situ XRD at 800 °C (XRD not shown), while γ-$Al_2O_3$ peaks are shifted to lower angles. The $V_2O_5$ peaks reappear in post-annealing room temperature XRD, visible in Figure 6 b). The loss of $V_2O_5$ intensity at 800 °C indicates liquefaction of a V-rich phase which then recrystallize as $V_2O_5$ upon cooling. This could explain the long-range diffusion needed in order to form the observed clusters, Figure 5. This liquefaction of a V-rich phase is according to expectation from the $Al_2O_3$-$V_2O_5$ pseudo binary phase diagram [31].

The observed crack pattern in Figure 5 for the 7 at.% V coating (sample G) after annealing to 900 °C could be either due to intrinsic coating effects or in-diffusion of substrate Si into the coating. To rule out the latter effect, a TEM lift out sample was made from one of the areas where the coating had not cracked and started to flake. The EDX-line scan over the entire coating thickness plus substrate showed no Si-gradient over the coating thickness and the level of Si was as low as Ar and V after annealing (see Figure S2, supplementary material). The evenly distributed Si-content in the coating can be explained by re-sputtering of Si from the substrate during FIB-preparation. The conclusion is therefore that the crack pattern originates from intrinsic coating stresses from when the V-leave the $Al_2O_3$ structure. Previous work on the nucleation of α-$Al_2O_3$ from a CVD grown κ-$Al_2O_3$ showed similar round crack pattern [54] and was then attributed to secondary phase nucleation in combination with volume contraction. In the here studied coating the secondary phase is such a small fraction, due to low V-content, that the possible grains attributed to the phase are only visible in SEM after annealing to 900 °C.

The measured nanohardness and reduced elastic modulus as a function of annealing temperature, Figure 7 a) and b) respectively, show no hardness increase with increased annealed temperature in conjunction with the additional phase formation of $V_2O_5$. This is possibly due to that the phase formed seem to originate at the surface of the as deposited coatings. The measured V-deficient Al$_2$O$_3$ coatings are then weaken by point defects and increased porosity, as visible from SEM (not shown). The mechanical properties of both the high oxidation state V-based phases, such V$_2$O$_5$, as well as the low oxidation state V$_2$O$_3$ (V$^{3+}$) are generally softer than α-Al$_2$O$_3$. For V$_2$O$_3$ and V$_2$O$_5$ it has been shown theoretically [29] and for V$_2$O$_3$ also in isostatic compression test [55].

The hardness reduction of the pure γ-Al$_2$O$_3$, sample H, after annealing to 1000 °C, Figure 7 a), is possibly due to grain growth, enhanced by the accidently prolonged annealing time of 3 h. There is no visible change in the LOM image after that annealing so the change would have to be on the submicroscopic level, such as slight grain growth and defect annihilation. The hardness loss and subsequent cracking of the pure γ-Al$_2$O$_3$ upon phase transformation to α-Al$_2$O$_3$ show the inferior mechanical properties of the gamma phase over the thermodynamically stable and denser α-Al$_2$O$_3$.

## 5. Conclusions

We have demonstrated reactive magnetron co-sputter deposited (Al$_{1-x}$V$_x$)$_2$O$_3$ coatings from elemental targets with x ranging from 0 to 1, deposited at 550 °C. The resulting crystal structures from XRD for the -120 V biased coating series can be divided into three different regions:

The pure vanadium oxide coating was synthesized in α-V$_2$O$_3$ rhombohedral structure. A defect spinel structure was obtained for the intermediate region, 63 - 42 at.% V metal fraction, with a possible two-phase structure for the 63 at.% V metal fraction sample consisting of a α-(V,Al)$_2$O$_3$ and a γ-like Al$_2$O$_3$. No coating with single phase α-(Al,V)$_2$O$_3$ solid solution was observed. For low V-metal fraction, 18 and 7 at.%, a gamma-alumina-like solid solution was observed, shifted to larger d-spacing compared to pure γ-Al$_2$O$_3$.

Annealing the Al-rich samples in air resulted in formation of $V_2O_5$ crystals on the surface of the coating after annealing to 500 °C for 42 at.% V and 700 °C for 18 at.% V metal fraction respectively. The highest thermal stability was shown for pure $\gamma$-$Al_2O_3$-coating, which transformed to $\alpha$-$Al_2O_3$ after annealing to 1100° C with associated coating cracking and flaking upon cooling.

The highest hardness was observed for the Al-rich oxide, ~24 GPa, and then it decreased with increasing V-content larger than 7 at.% V metal fraction. The 7 at.% V in the $Al_2O_3$ coating resulted in a significant surface smoothening compared to the binary oxide. The measured hardness after annealing in air decreased in conjunction with the onset of further oxidation of the coatings and no age hardening was observed.

This work increases the understanding of this complex material system with respect to phase formation and their response to annealing in air. Further optimization to the coating system, focusing on, optimized stress levels, binding layer and substrate combinations, may mitigate the cracking behavior and make it possible to benefit from the low friction properties of V-based oxides.

6. Acknowledgments

The Swedish Research Council (VR, grant number: 621-212-4368) is acknowledged for financial support for L. L.'s industry PhD studies with AB Sandvik Coromant. P. E. acknowledges the Knut and Alice Wallenberg Foundation through a Fellowship grant, and the Swedish Government Strategic Research Area in Materials Science on Functional Materials at Linköping University (Faculty Grant SFO-Mat-LiU No. 2009 00971. B.A. acknowledge financial support from the Swedish Research Council (VR) through the International Career

Grant No. 330-2014-6336 and by Marie Sklodowska Curie Actions, Cofund, Project INCA 600398, as well as from the Swedish Foundation for Strategic Research (SSF) through the Future Research Leaders 6 program. Support by VR-RFI (contracts #821-2012-5144 & #2017-00646_9) and the Swedish Foundation for Strategic Research (SSF, contract RIF14-0053) supporting accelerator operation at Uppsala University is gratefully acknowledged. M. Moro is acknowledged for help with ERDA and RBS interpretation and fruitful discussions.

Supplementary material for

# Phase composition and transformations in magnetron-sputtered (Al,V)$_2$O$_3$ coatings

L. Landälv[*,a,b], C.-F. Carlström[b], J. Lu[a], D. Primetzhofer[c], M. J. Jõesaar[d], M. Ahlgren[b], E. Göthelid[b], B. Alling[e], L. Hultman[a], P. Eklund[*,a]

[a] Thin Film Physics Division, Department of Physics, Chemistry, and Biology (IFM), Linköping University, Linköping SE-581 83, Sweden
[b] Sandvik Coromant AB, Stockholm SE-126 80, Sweden
[c] Applied Nuclear Physics, Department of Physics and Astronomy, Uppsala University, Uppsala SE-751 20, Sweden
[d] SECO Tools AB, SE-737 82 Fagersta, Sweden
[e] Theoretical physics, Department of Physics, Chemistry, and Biology (IFM), Linköping University, Linköping SE-581 83, Sweden

**Table S1.** Unit cell fit with program "unit cell". The stated lattice spacings are based on extracted Kα1 values from fitted peaks using "Highscore", instrumental broadening has not been considered.

| With 222, 400, 333, 440, 444 planes included | $a$ (Å) | $c$ (Å) | Tetragonal Unit Cell volume (Å$^3$) |
|---|---|---|---|
| Al$_2$O$_3$ | 7.92±0.01 | 7.90±0.01 | 495.5±0.1 |
| Al$_2$O$_3$, 7 at.% V metal fraction | 7.96±0.01 | 7.95±0.01 | 503.7±0.1 |
| Al$_2$O$_3$, 18 at.% V metal fraction | 7.99±0.01 | 7.98±0.01 | 509.5±0.1 |

*Corresponding authors
Email address: ludvig.landalv@liu.se (L. Landälv)
Per.eklund@liu.se (P. Eklund)

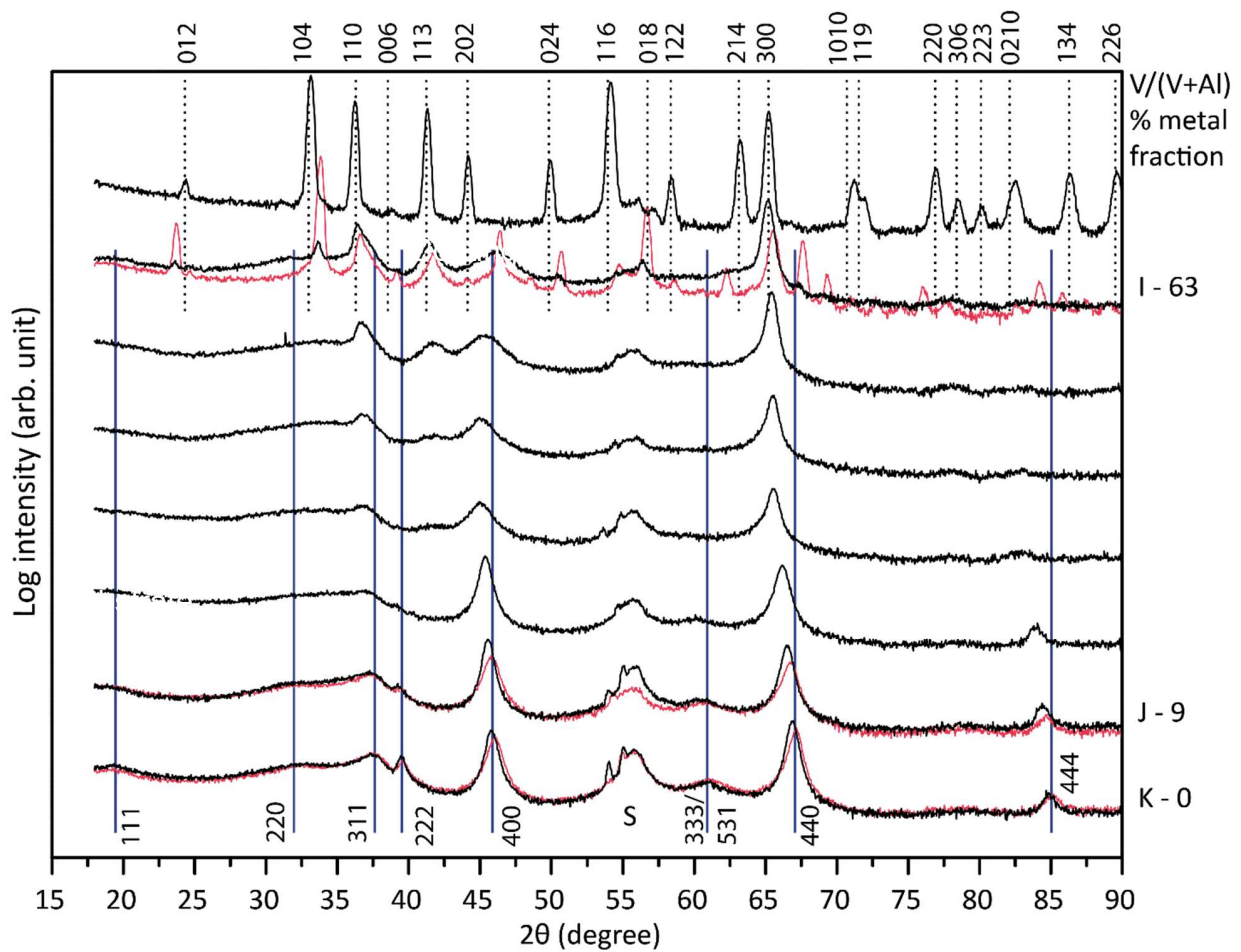

**Figure S1.** XRD-diffractogram for coatings synthesized with -120 V bias with overlayered coatings deposited with -20 V bias (marked in red). Decreasing V-content from the top. "S" marking Si-substrate (311), blue lines γ–Al$_2$O$_3$, long dots defect spinell-AlVO$_3$ and fine dotted lines α-V$_2$O$_3$. Grazing incidence XRD with 1.5° incoming beam and parallel plate collimator plus open detector, Cu K$_α$.

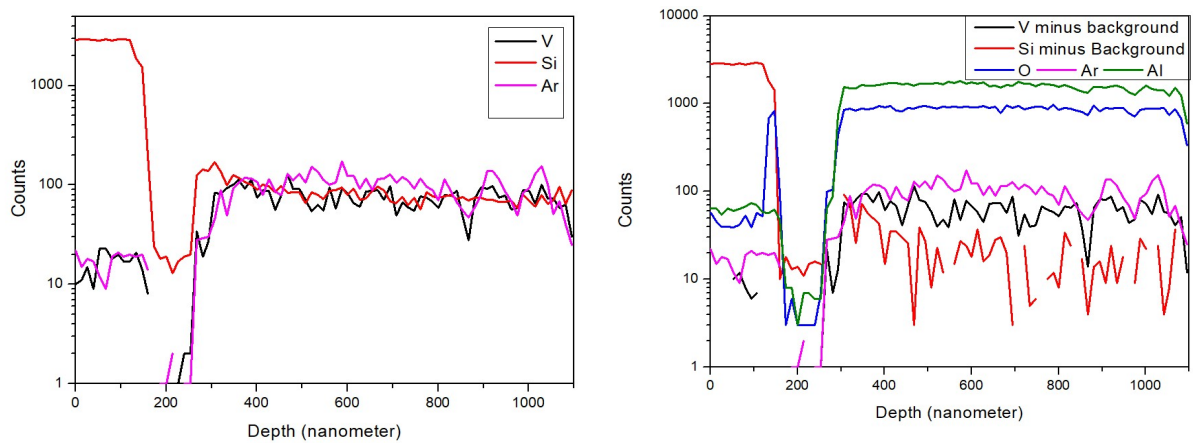

**Figure S2.** See figure caption on next page

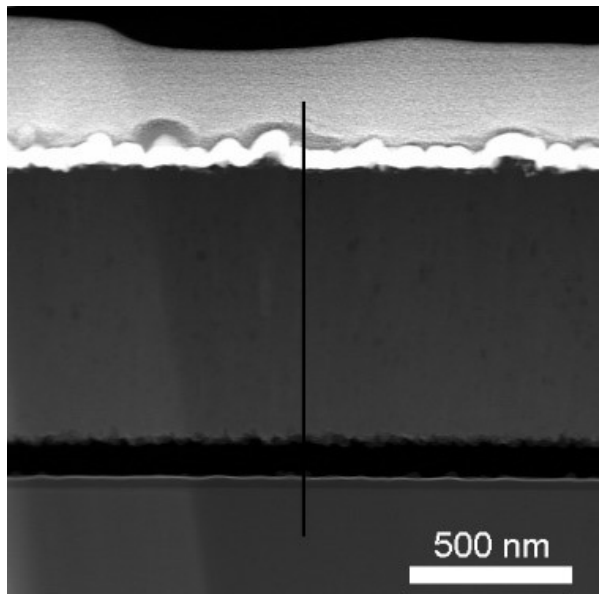

**Figure S2.** FIB+TEM on 8 at.%V mf sample to understand if Si has diffused into the sample. Line-profiles (bottom) along black line in HAAF-micrograph (top) with Si-substrate from the left in the depth-profiles. Constant Si-content on a very low level (close to background noise level) over the entire thickness, hence no Si-in-diffusion into the coating.